\documentclass[twoside,twocolumn,9pt]{article}
\usepackage{extsizes}
\usepackage[super,sort&compress,comma]{natbib} 
\usepackage[version=3]{mhchem}
\usepackage[left=1.5cm, right=1.5cm, top=1.785cm, bottom=2.0cm]{geometry}
\usepackage{balance}
\usepackage{mathptmx}
\usepackage{sectsty}
\usepackage{graphicx} 
\usepackage{lastpage}
\usepackage[format=plain,justification=justified,singlelinecheck=false,font={stretch=1.125,small,sf},labelfont=bf,labelsep=space]{caption}
\usepackage{float}
\usepackage{fancyhdr}
\usepackage{fnpos}
\usepackage[english]{babel}
\addto{\captionsenglish}{%
  
}
\usepackage{array}
\usepackage{droidsans}
\usepackage{charter}
\usepackage[T1]{fontenc}
\usepackage[usenames,dvipsnames]{xcolor}
\usepackage{setspace}
\usepackage[compact]{titlesec}
\usepackage{hyperref}

\usepackage{orcidlink}

\usepackage{epstopdf}
\usepackage{stfloats}

\definecolor{cream}{RGB}{222,217,201}

\begin{document}

\pagestyle{fancy}
\thispagestyle{plain}
\fancypagestyle{plain}{
\renewcommand{\headrulewidth}{0pt}
}

\makeFNbottom
\makeatletter
\renewcommand\LARGE{\@setfontsize\LARGE{15pt}{17}}
\renewcommand\Large{\@setfontsize\Large{12pt}{14}}
\renewcommand\large{\@setfontsize\large{10pt}{12}}
\renewcommand\footnotesize{\@setfontsize\footnotesize{7pt}{10}}
\renewcommand\scriptsize{\@setfontsize\scriptsize{7pt}{7}}
\makeatother

\renewcommand{\thefootnote}{\fnsymbol{footnote}}
\renewcommand\footnoterule{\vspace*{1pt}%
\color{cream}\hrule width 3.5in height 0.4pt \color{black}\vspace*{5pt}} 
\setcounter{secnumdepth}{5}

\makeatletter 
\renewcommand\@biblabel[1]{#1}            
\renewcommand\@makefntext[1]%
{\noindent\makebox[0pt][r]{\@thefnmark\,}#1}
\makeatother 
\renewcommand{\figurename}{\small{Fig.}~}
\sectionfont{\sffamily\Large}
\subsectionfont{\normalsize}
\subsubsectionfont{\bf}
\setstretch{1.125} 
\setlength{\skip\footins}{0.8cm}
\setlength{\footnotesep}{0.25cm}
\setlength{\jot}{10pt}
\titlespacing*{\section}{0pt}{4pt}{4pt}
\titlespacing*{\subsection}{0pt}{15pt}{1pt}

\fancyfoot{}
\fancyfoot[RO]{\footnotesize{\sffamily{\thepage}}}
\fancyfoot[LE]{\footnotesize{\sffamily{\thepage}}}
\fancyhead{}
\renewcommand{\headrulewidth}{0pt}
\renewcommand{\footrulewidth}{0pt}
\setlength{\arrayrulewidth}{1pt}
\setlength{\columnsep}{6.5mm}
\setlength\bibsep{1pt}

\makeatletter 
\newlength{\figrulesep} 
\setlength{\figrulesep}{0.5\textfloatsep} 

\newcommand{\topfigrule}{\vspace*{-1pt}%
\noindent{\color{cream}\rule[-\figrulesep]{\columnwidth}{1.5pt}} }

\newcommand{\botfigrule}{\vspace*{-2pt}%
\noindent{\color{cream}\rule[\figrulesep]{\columnwidth}{1.5pt}} }

\newcommand{\dblfigrule}{\vspace*{-1pt}%
\noindent{\color{cream}\rule[-\figrulesep]{\textwidth}{1.5pt}} }

\makeatother

\graphicspath{{Figures}}

\twocolumn[
  \begin{@twocolumnfalse}
\vspace{1em}
\sffamily
\noindent\LARGE{\textbf{Energy-Optimal Allocation of Storage in Transmission Grid Networks$^\dag$}}

\vspace{0.4cm}
\noindent\large{Emile Emery\orcidlink{0000-0001-5074-1163},\textit{$^{a}$}$^{\ast}$ Sébastien Aumaître,\textit{$^{a}$} and Hervé Bercegol\orcidlink{0000-0002-1792-7756}\textit{$^{a}$}}

\vspace{0.4cm}
\noindent\normalsize{\textbf{Abstract:} The deployment of renewable energy technologies supposes the connection to the power grid of many new, distributed, and variable electricity production facilities. Among the investments deeply needed for a successful shift to clean energy, electricity storage systems are key to provide power reliably, continuously and economically. Here, we are concerned with the energy that must be invested and embodied in storage devices and in production oversizing to cope with natural variations of renewable electricity production, and compensate for any gap between production and consumption. We developed a model to analyze the variation of energy expenses with the location in the grid, capacity of storage and production oversizing. We apply it to a time scale of fluctuations of a few hours that can be taken care of by Li-ion batteries to calculate the optimal storage capacity and production oversizing. It yields a maximum value of the ESOI ratio [Energy Stored On energy Invested] at a given satisfaction rate of customer demand. We evaluate these values for a rescaled present-time French power mix  and two idealized zero-emission mixes (100\% PV and 100\% wind). In parallel, using a recently developed model of French transmission grid, a centrality-based analysis shows that locating storage at nodes of maximal installed power minimizes additional Joule losses. These results generalize existing grid-level energy return frameworks to incorporate storage sizing, location, and transmission losses into a unified assessment of future power grid configurations.}

 \end{@twocolumnfalse} \vspace{0.6cm}

  ]

\renewcommand*\rmdefault{bch}\normalfont\upshape
\rmfamily
\section*{}
\vspace{-1cm}

\footnotetext{\textit{$^{a}$~Université Paris-Saclay, CEA, CNRS, SPEC, 91191, Gif-sur-Yvette, France }}
\footnotetext{\textit{$^{\ast}$~E-mail: emile.emery@etik.com }}







\section{Introduction}

The shift toward renewable energy relies on expanding distributed electricity generation, but this brings uncertainty, voltage-fluctuation and reliability problems. Coordinated deployment of distributed generation together with energy‑storage systems (ESS) can alleviate many of these issues \cite{zhang_systematic_2022}. The capacity to store a fraction of the produced energy during over-production periods to release it during over-consumption is a powerful flexibility lever. 

ESS rely on a variety of physical processes—mechanical, thermal, magnetic, chemical, or electrochemical—each with its own advantages and drawbacks \cite{yao_challenges_2016, sgouridis_comparative_2019}. On the one hand, building an ESS requires embodied energy, related to an unavoidable environmental impact. On the other hand, excess generation that cannot be stored is often deliberately wasted to maintain stability in a process called curtailment. Deciding whether to store or to curtail energy is essentially an energy‑return‑on‑investment (EROI) problem. Barnhart and Benson therefore introduced the Energy Stored on Invested (ESOI) metric, which compares the lifetime amount of energy stored to the energy expended to build the device, indicating whether to store or to curtail is the best energy-saving strategy \cite{barnhart_importance_2013, barnhart_energetic_2013}. Pellow \textit{et al.} \cite{pellow_hydrogen_2015} expanded the global EROI framework to incorporate storage, creating a grid‑level EROI that balances three factors: the proportion of over‑generated energy $\varphi$, the round‑trip efficiency of the storage $\eta_\mathrm{st}$, and the relative EROIs of generation and storage:
\begin{equation}\label{pelloweroigrid}
    \text{EROI}_\text{grid} = \frac{1 - \varphi + \eta_{\mathrm{st}}  \varphi}{\frac{1}{\text{EROI}_{\text{gen}}} + \frac{\varphi}{\text{ESOI}}}.
\end{equation}
This formulation adds information on losses through $\eta_\mathrm{st}$ only. Clerjon and Perdu later refined the ESOI concept by adding the extra energy needed to over‑size generation to compensate for storage losses (because storage efficiency is never perfect). Their generalized ESOI therefore accounts for both the primary embodied energy of the storage device and the additional production required to compensate imperfect efficiencies \cite{clerjon_matching_2019}. Using Haar wavelet decomposition \cite{haar_zur_1910}, they also broke down the contribution of storage across different timescales, allowing a nuanced view of how various storage technologies perform over various fluctuation durations. In this landscape, where storage should be placed within the power grid network also matters and remains an open question \cite{rafaq_comprehensive_2022}. Clerjon \textit{et al.} recognized that storage location and Joule losses were blind spots of their work \cite{clerjon_matching_2019}. Korjani \textit{et al.} showed that optimal positioning can cut line losses by roughly ten percent and improve voltage stability, when the storage nodes occupy minimal eigenvector centrality sites \cite{korjani_aging_2017,korjani_optimal_2018}. 
Like most of the ESS location optimization studies, these works only focus on the distribution network \cite{mohamad_optimum_2021, bhattacharya_optimising_2024} with methods that may not be applicable to the transmission level. Although some field tests have been conducted such as the Ringo project of the French transmission grid operator RTE \cite{gionfra_balanced_2023}, the high voltage transmission part remains understudied. To study scenarios where investment in storage system would decrease the need for an increased production, we must build models to determine the associated optimal storage strategy. This strategy is composed in two parts: we want to know how much capacity we should have and how to distribute it spatially. Answering both questions could give important insights for the assessment of future energy mixes.

We developed a novel modelling of storage dynamics that takes as input production and consumption signals based on governmental time series. Then, we use the ESOI indicator as a tool to identify and quantify the various energy investments involved and evaluate the trade-offs. Afterward, given an instantaneous production layout and a topology, applying centrality measures from graph theory gives informed characterizations of nodes and relevant estimates of the Joule losses over a transmission power grid \cite{emery_complex_2024,emery_power_2025}. 

This paper proposes a new methodology to evaluate storage strategies at the country-level production, and to consider losses in-line in the global energy analysis of an electricity grid system with production, consumption, and storage.

\section{Methods}

\subsection{Treatment of the production signal}

\begin{figure}[t]
        \centering
        \includegraphics[width=1\linewidth]{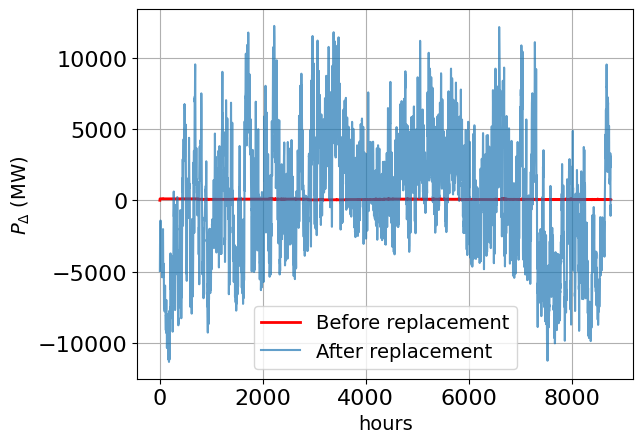}
        \caption{Variation over the year 2021 of the power discrepancy. In the raw data (in red), production follows consumption; consequently discrepancy is zero. We are interested in the blue signal, obtained by removing carbonized production and multiplying renewable production by $\alpha$. This paper determines how to cope with short time discrepancies, by adding an optimal capacity of Li-ion batteries and power oversizing.}
        \label{fig:P_delta}
\end{figure}

The production signal is the sum of the contributions from nuclear ($N$), solar ($S$), wind ($W$), hydro ($H$), bio‑energy ($B$), oil ($O$), coal ($C$), and gas ($G$). The consumption signal $C(t)$ is the sum of an effective demand term, a pumping term, and an international exchange term. In the raw dataset, the two signals are equal at every hourly time step, so no storage is required. To create a realistic storage need, we modify the production by replacing the fossil-fuel based generators with a rescaling of wind and solar. The production after replacement (or "rescaled production") is
\begin{equation} \label{replacement}{P}(t)=\alpha\Bigl(P_{\mathrm{S}}(t)+P_{\mathrm{W}}(t)\Bigr)+P_{\mathrm{N}}(t)+P_{\mathrm{H}}(t)+P_{\mathrm{B}}(t), \end{equation}
where the scaling factor $\alpha$ is chosen so that the total annual energy from the original carbon sources equals the additional renewable energy. 
%
%
A key quantity for the storage‑flux analysis is the hourly power discrepancy between (rescaled) production and consumption $P_\Delta(t)=P(t)-{C}(t)$. As pictured in Figure \ref{fig:P_delta}, this signal is constant before the rescaling, but the rescaling of renewable production makes it non-trivial. Integrating it over a given interval yields the amount of curtailed energy that must be supplied to keep consumption satisfied over this interval. In the initial production signal, the oil, gas, and coal are used more in winter. This makes $ P_\Delta(t)$ largely negative in winter (excess demand) and positive in summer (excess supply), leading to a multi‑terawatt‑hour storage requirement (for France), far beyond the gigawatt‑hour scale of existing non‑hydraulic storage \cite{ODReinstallations}. In practice, inter‑seasonal balancing is achieved by dispatchable generation (including nuclear which is considered flexible at this timescale) and large‑scale hydro reservoirs, while pumped‑storage hydropower addresses week‑to‑month fluctuations. At shorter timescales, lithium-ion battery storage — whose global installed capacity has grown from about 1 GW in 2013 to over 85 GW in 2023, driven by steep cost reductions~\cite{IEA_batteries_2024} — increasingly handles day-to-day and intra-day fluctuations~\cite{clerjon_matching_2019}. Looking ahead, emerging long-duration storage technologies — in particular power-to-hydrogen — may eventually complement hydro reservoirs for seasonal balancing~\cite{dowling_role_2020}.

The Haar wavelet decomposition provides a complete orthogonal function system based on piecewise-constant functions \cite{haar_zur_1910}. The observation period is recursively divided: at resolution level $j$, the time axis is divided into $2^j$ equal sub-intervals, and the corresponding Haar basis functions take opposite constant values on pairs of adjacent sub-intervals and zero elsewhere. Any square-integrable signal can be decomposed into contributions at each timescale. We denote by $\Omega_\tau$ the subset of resolution levels retained for the analysis. 
To model storage dynamics at this specific timescale, we will use the Haar wavelet decomposition to build a filtered production signal that can be written as 
\begin{equation}\label{prod_haar} \tilde P(t,\Omega_\tau)=\langle P\rangle+H_{P}(t, \Omega_\tau), 
\end{equation}
where $H_{P}(t,\Omega_\tau)$ is the Haar‑reconstructed component of the rescaled production $P$, with $\Omega_\tau$ the set of timescales under consideration. $\langle P\rangle$ is the mean value (on the considered duration) of the rescaled production, it is equal to the mean of the original production and consumption series. Similarly, the consumption signal becomes 
\begin{equation} \label{conso_haar}\tilde C(t,\Omega_\tau)=\langle P\rangle+H_{C}(t,\Omega_\tau), \end{equation}
with $H_{C}(t,\Omega_\tau)$ the Haar‑reconstructed component of the original consumption $C(t)$. Note that by construction of the rescaling (Eq. \ref{replacement}), the annual mean production equals the annual mean consumption, so the same constant appears in both Eqs. \ref{prod_haar} and \ref{conso_haar}. From these definitions, we obtain a Haar reconstructed power discrepancy measure $  \tilde P_\Delta(t,\Omega_\tau)=\tilde P(t,\Omega_\tau)-\tilde C(t,\Omega_\tau)$.

\subsection{Dynamical storage model}

\begin{figure}[t]
        \centering
        \includegraphics[width=1\linewidth]{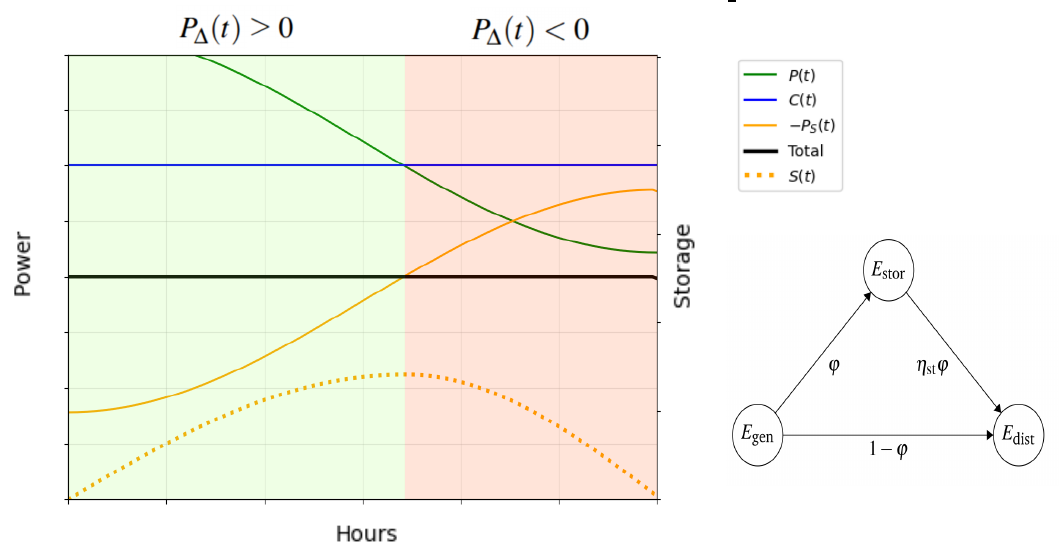}
        \caption{Example of a simple power evolution obtained by the dynamical storage model. Here we take $P(t) = P_0  (1+\sin(\omega t)) + P_1$ and $C(t) = P_0$. We have $P_0$ as the instantaneous average power produced and consumed, and $P_1$ is the oversizing that is needed to compensate the fact that $\eta_{\mathrm{st}}$ is lower than 1. The total value correspond to the sum $P(t)+P_{\mathrm{st}}(t)-C(t)$. \textbf{Bottom right:} Schematic picture of an energy system including storage.}
        \label{fig:trianglescheme}
\end{figure}

In the following, we assume that we have fixed $\Omega_\tau$ and omit mentioning it as a parameter. We set $E_\text{gen}$,  $E_\text{stor}$ and $E_\text{dist}$ to represent respectively the generated, the stored and the distributed energy on a given duration. The relationships between these energy slots are such that $E_\text{stor} = \varphi E_\text{gen}$ and $E_\text{dist}= \eta_{\mathrm{st}}\varphi E_\text{gen} + (1-\varphi) E_\text{gen}$, where $\varphi \in [0,1]$ is the fraction of generated energy routed to storage. This simplified storage dynamics is represented by the triangular scheme in Figure \ref{fig:trianglescheme}. The model iterates over each time step as follows: in case of overproduction, the extra power is stored in the ESS; in case of underproduction, the stored energy is consumed. However, extra constraints must be added to ensure the feasibility of the model. If the ESS is fully charged at an overproduction step, power curtailment occurs with an associated loss in efficiency. If the ESS is empty during an underproduction step, then consumption is not satisfied at this step. This case needs to be avoided as much as possible to have a viable system. In this purpose, we will later add a production power $P_1$ called \textit{oversizing}. Let $S(t)$ denote the stored energy at time $t$. Its evolution over a time step $\Delta t$ is


\begin{equation} \label{storagestock}
   S(t +\Delta t)= \left\{
\begin{array}{ll}
S(t) +\Delta t \;P_{\mathrm{st}}(t)  & \text{if } \tilde P_\Delta(t)\geq 0 \\
S(t) +\frac{1}{\eta_{\mathrm{st}}}\;\Delta t \;P_{\mathrm{st}}(t)& \text{if } \tilde P_\Delta(t)<0,
\end{array}
\right.
\end{equation}
where $\eta_{\mathrm{st}}$ is the round‑trip efficiency, and $P_{S}(t)$ is the power exchanged with the grid which is defined as:
\begin{equation} \label{storageflux}
    P_{\mathrm{st}}(t) = \left\{
\begin{array}{ll}
\min\Big( \tilde P_\Delta(t)\;,\; \frac{S_{\text{max}} -S(t)}{\Delta t},\;P_\text{max}\Big) & \text{if } \tilde P_\Delta(t)\geq0 \\
\max\Big(\tilde P_\Delta(t)\;,\; -\frac{\eta_{\mathrm{st}}S(t)}{\Delta t}, -\;P_\text{max}\Big) & \text{if } \tilde P_\Delta(t)<0,
\end{array}
\right.
\end{equation}
with $P_\text{max}$ the maximal power emitted or absorbed by the grid-level ESS, and $S_\text{max}$ the maximum capacity of the complete storage system. The power input/output coming to/from storage is limited by three terms: the instantaneous surplus or deficit, the remaining storage capacity, and the maximum power of the storage system.\\

\begin{figure}[t]
        \centering
        \includegraphics[width=1\linewidth]{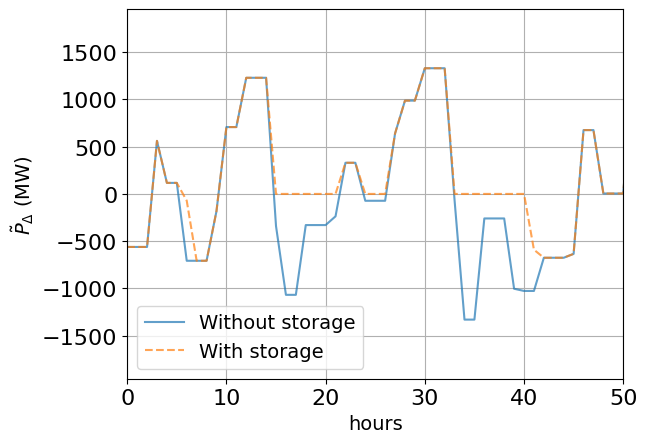}
        \caption{Variation over 50 hours of the Haar reconstructed power discrepancy in blue. The orange signal is obtained by adding the storage power to $\tilde P_\Delta(t)$.}
        \label{fig:P_delta_tilde}
\end{figure}

Using the storage dynamics model, we can estimate the stored energy $S(t)$ at each hour $t$ ($\Delta t=1$h). As shown in Figure \ref{fig:P_delta_tilde}, the storage dynamics stabilize $ \tilde P_\Delta(t)$ 
toward zero during many intervals that would otherwise be negative, but residual deficits remain because (i) $\eta_{\mathrm{st}}<1$ leading to a known need for oversizing \cite{clerjon_matching_2019} and (ii) the storage may not have sufficient time to charge before entering an over‑consumption period (the previous over-production period was not sufficient). Increasing storage capacity alone does not resolve all the deficit, this is where the oversizing $P_1$ comes in.

\subsection{Energy Stored On energy Invested}

In this paper, the ESOI metric is defined as the ratio of the following energies (all refer to a unit time duration):

\begin{equation}
    \text{ESOI} = \frac{E_U^s+E_U^o}{E_I^s+E_I^o}
\end{equation}
where the $U$ subscript stands for \textit{useful} and $I$ for \textit{investment}. The $s$ superscript refers to \textit{Storage} and $o$ to \textit{Oversizing}. So for a given storage-oversizing pair ($S_\text{max}$, $P_1$) we can estimate the ESOI value by running the model over a given time range. Joule losses are neglected for now. For a given strategy, $E_U^s$ equals $ \eta_\text{st} \varphi E_\text{gen}$ and $E_U^o$ corresponds to remaining energy deficits that have been filled by oversizing. The ESS energy investment is given by\cite{clerjon_matching_2019}:

\begin{equation}
    E_I^s = \frac{\max(\zeta_E  S_\text{max}, \zeta_P  P_\text{max})}{\min( \tau, N_c^\text{max}/N_c)},
\end{equation}
where $\zeta_E$ and $\zeta_P$ are the energy intensity in energy and power for the ESS, $\tau$ is the lifetime of the ESS, $N_c$ is total number of charging/discharging cycles that occur during the timerange and $N_c^\text{max}$ is the maximum number of cycles. To estimate the energy invested in oversizing, we estimate first the energy investment of the production system without oversizing. For this, we use energy intensity normalized by the power installed, taking values from Clerjon \textit{et al.}\cite{clerjon_matching_2019}. To simplify, oversized production will be a slight upscaling of the current production system, keeping energy mix proportion unchanged. Then, the energy investment of oversizing $E_I^o$ is simply the corresponding fraction of the energy investment of the production system without oversizing. 

\subsection{Centrality analysis}

When topology is taken into account, the additional Joule losses caused by the extra power‑flow paths to and from storage must be considered. In this regard, we used previously developed methods\cite{emery_complex_2024,emery_power_2025} that take as input a weighted power grid network with information of the algebraic energy production on each node as well as voltage and length of each edge, and gives as output an estimation of the total Joule losses through DC approximation\cite{mccalley_power_nodate}. To distribute power among the nodes, we assign to each node $n$ at time $t$ the following production weight:
\begin{equation} p_{n}(t)=\beta(t)\Bigl( \varepsilon^{\mathrm{W}}_{n}\langle P_{\mathrm{W}}\rangle+\varepsilon^{\mathrm{S}}_{n}\langle P_{\mathrm{S}}\rangle  \Bigr)+\varepsilon^{\mathrm{N}}_{n}\langle P_{\mathrm{N}}\rangle +\varepsilon^{\mathrm{H}}_{n}\langle P_{\mathrm{H}}\rangle +\varepsilon^{\mathrm{B}}_{n}\langle P_{\mathrm{B}}\rangle, \end{equation}
where the scaling factor is defined as
\begin{equation} \beta(t)=\frac{H_{P}(t)+\langle P_{\mathrm{C}}\rangle+\langle P_{\mathrm{O}}\rangle+\langle P_{\mathrm{G}}\rangle} {\langle P_{\mathrm{W}}\rangle+\langle P_{\mathrm{S}}\rangle}+1, \end{equation}
$\langle P_i\rangle$ is the average production over the year made by type $i$, and the factors $\varepsilon^{i}_n$ are the fraction of the annual energy produced by type $i$ at node $n$. They weigh the attribution of the global production signal to individual nodes. By construction, $\sum_{n}p_{n}(t)=\tilde P(t)$ as required. We divide departmental consumption value by the number of substations in that department and distribute uniformly among the nodes in that department. This is done using governmental databases about departmental consumption \cite{conso} and France's departmental divisions \cite{map}. Correspondingly, in the presence of storage, we attribute an additional power weight $P_{\mathrm{st}}(t)/N_S$ to all the $N_S$ storage nodes. 


\section{Results}



\begin{figure*}[p]
    \centering
    \includegraphics[width=\linewidth]{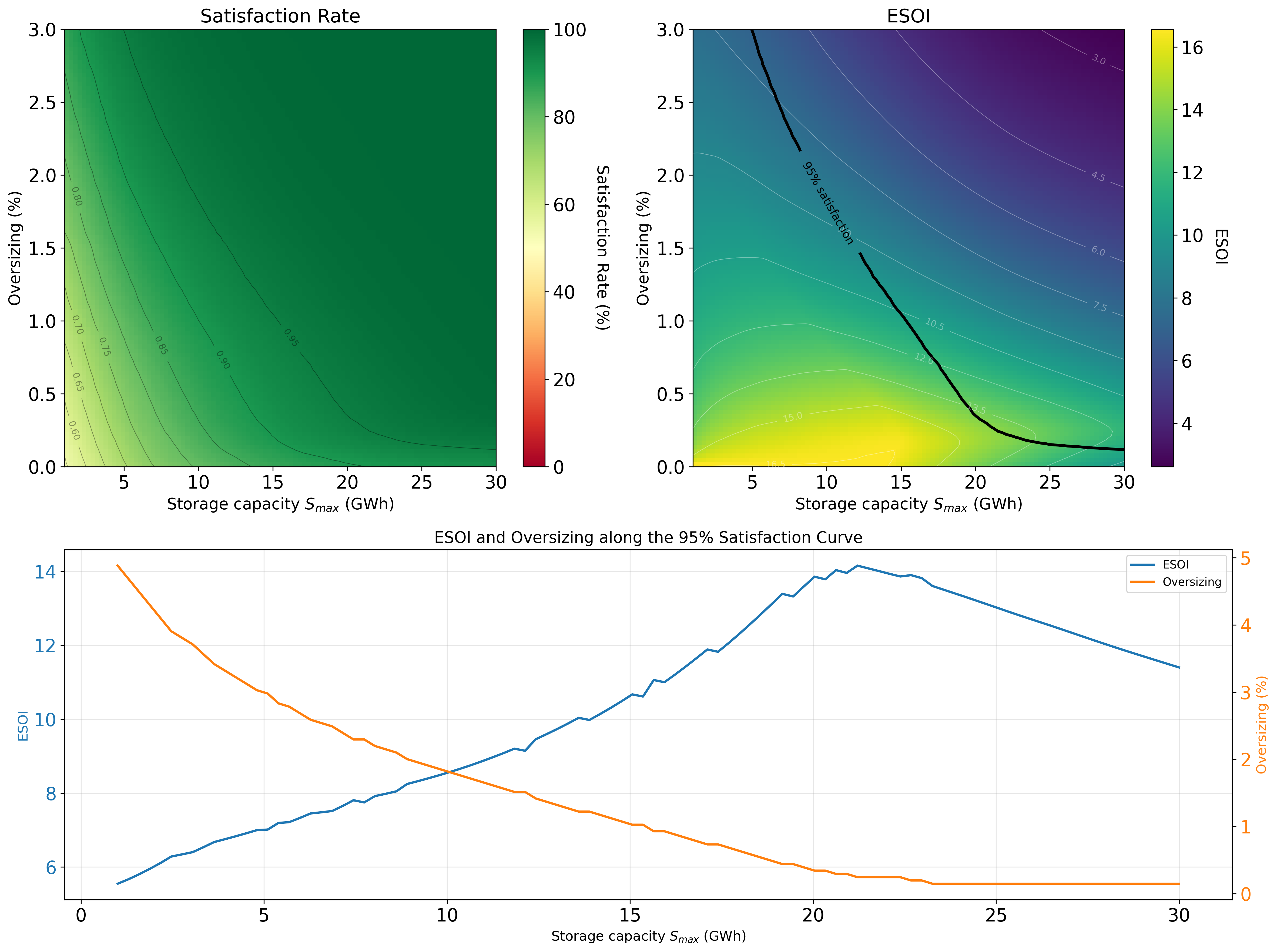}
    \caption{Characterizations of 10000 iterations of the model for 100 values of $S_\text{max}$ and 100 values of $P_1$. \textbf{Upper left:} Values of satisfaction rate. \textbf{Upper right:} Values of ESOI. \textbf{Lower panel:} Projection of the ESOI values on the subset of parameters corresponding the 95\% satisfaction rate.}
    \label{fig:Smax}
\end{figure*}
\begin{figure*}[p]
    \centering
    \includegraphics[width=0.8\linewidth]{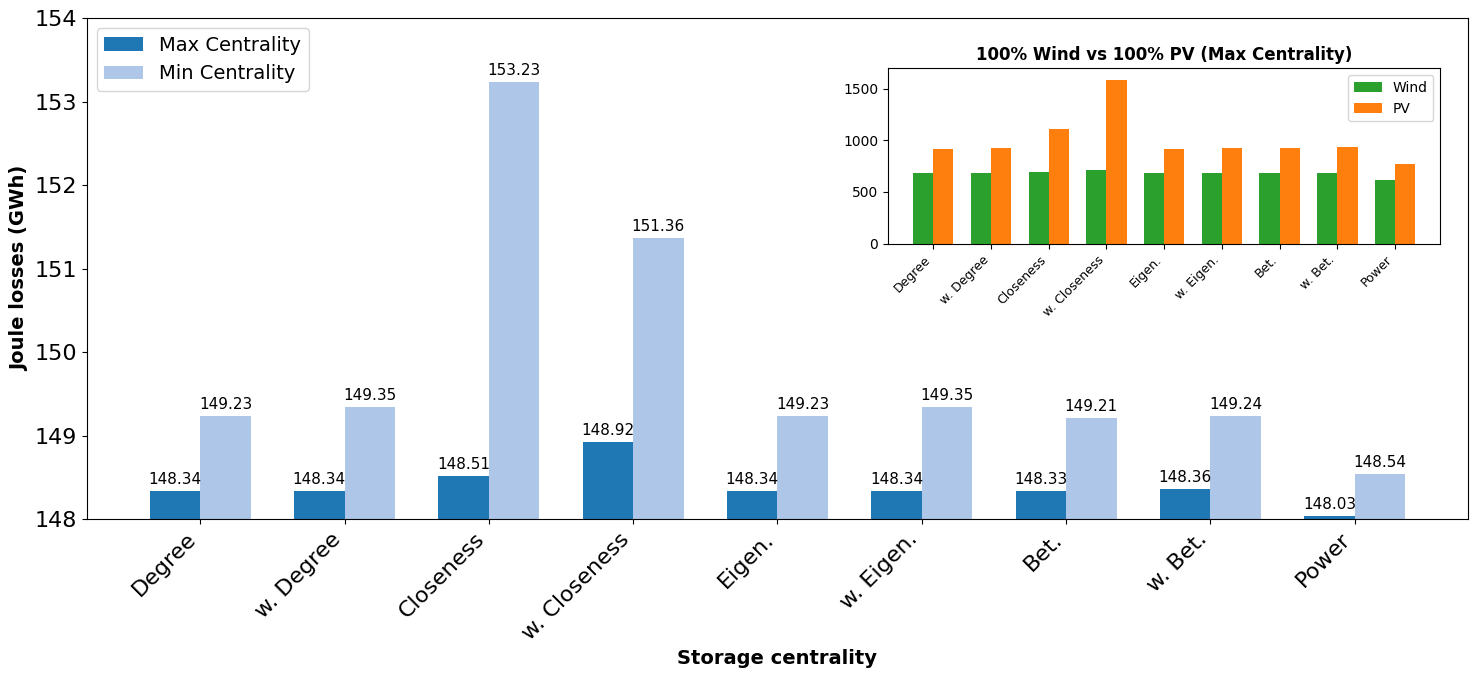}
    \caption{Total Joule losses for various scenarios and centrality functions, integrated over 250 hours. The power centrality corresponds to the production ranking of the nodes. The main figure represents the rescaled "RTE mix" case and how localizing storage on centrality extremes affects a few percents of the total losses. The inset figure focuses on the maximum centrality only but with 100\% PV and 100\% Wind energy mixes. Here, there is a high impact of storage location on losses. 100\% renewables have higher losses than present RTE mix, and inadequate localization can double this effect.}
    \label{fig:centrality}
\end{figure*}




We focus our analysis on the Li-ion battery as ESS, on French transmission grid and use the 2021 production and consumption data \cite{ODReenergymix}. Thus, we take the values from Clerjon and Perdu as parameters of the components, and select the 6–12h time window, which matches the typical operation timescale for battery storage. To estimate the optimal storage capacity, we ran the model over the year and compute the $\text{ESOI}$ value for a set of $(S_\text{max}, P_1)$ pairs. The optimal pair is the one that maximizes the $\text{ESOI}$ while keeping consumption satisfied most of the time. To quantify it, we define the satisfaction rate as the number of time step where production is greater or equal than consumption divided by the total number of time steps. We want a reasonable satisfaction rate of at least 95\%. We do not need to satisfy a non-realistic 100\% rate since our model does not aim to implement extreme event compensation strategies but evaluate steady state regime. To go from 95\% to 100\%, we should upgrade the model to consider inter-countries power exchanges for example. The Figure \ref{fig:Smax} show the values of ESOI and satisfaction rate for a set of parameters. We can deduce from it the optimal storage capacity and oversizing. In an "RTE mix" where all the carbon-based production has been replaced by a rescaling of PV and Wind energy, the best setup is to add 21.21 GWh of battery storage and oversize the production by 0.24\%. The figures in appendix show results equivalent to Figure \ref{fig:Smax} for extreme scenarios: 100\% PV and 100\% Wind. Even though those streamlined scenarios may seem unrealistic, it helps to understand the boundary conditions of the problem. The optimal parameters and the associated ESOI value are summarized in Table \ref{tab:FrenchIndicatorComp}.\\

\begin{table}[t]
    \centering
    \begin{tabular}{|c|c|c|c|}
    \hline
        &RTE rescaled & 100\%Wind &100\%PV \\
        \hline
        $\text{ESOI}_\text{max}$ & 14.16 & 12.31 & 9.84 \\
        \hline
        $S_\text{max}$ & 21.21 GWh & 142.95 GWh  & 921.44 GWh\\
        \hline
        Oversizing & 0.24\% &1.95\% &  8.20\%\\
        \hline
    \end{tabular}
    \caption{Optimal storage capacity, oversizing, and the corresponding ESOI value for various scenarios.}
    \label{tab:FrenchIndicatorComp}
\end{table}

Given the optimal value of storage capacity and oversizing, we can run the storage dynamics over a network topology, distributing the production, consumption, and storage power over the nodes at each time step. We allocate the $N_S$ storage site at nodes which extremise a given centrality measure and estimate the corresponding aggregated Joule losses. We choose $N_S = 50$ so that storage is concentrated on a well-defined subset of nodes, making the location strategy meaningful. In the limit $N_S \sim N$, nearly every node carries storage and the spatial allocation becomes irrelevant: the network topology no longer discriminates between location strategies. Applying this protocol to various centralities and energy mixes leads to various values of additional Joule losses given in Figure \ref{fig:centrality}. We remark on this figure that the optimal strategy is to attribute storage at the node of maximal installed power. In this way, during over-production, the extra power is going to be stored directly instead of flowing through the overcharged edges. Since the Joule losses depend directly on the square of the power flow in the DC approximation made here\cite{emery_complex_2024,emery_power_2025}, it is better to select a storage location in order to minimize the flow at each edge. Also, in a hierarchical top-down feeding grid like the current RTE network \cite{emery_complex_2024}, the storage nodes will benefit from the shortcut topology of the very high-voltage layer.

Storage dynamics modifies global power flows. To verify if the centrality strategy does not lead to congestion in the network, we verify that the condition 
\begin{equation}
    P_{ij} < \sqrt 3 {U_{ij} I_{\text{max}}  \cos(\phi)}
\end{equation}
is met for all edges and all centrality strategies for the rescaled production. Here, $P_{ij}$ is the power flow and $U_{ij}$ the voltage on edge $(i,j)$.  $ I_{\text{max}}= 2$kA is estimated from the literature \cite{pays1994cables}. The power factor value is $cos(\phi) = 0.9$, which is a standard approximation regarding transmission network.

\section{Discussion \& Conclusion}

In this paper, we have shown that it is possible to find an energy optimal way to solve the consumption/production discrepancy in a renewable dominated mix. For a given timescale, at a set percentage of demand satisfaction, a Haar decomposition allows to determine the best way to invest energy into the storage system, including the necessary oversizing of production. The optimal storage capacity varies with the nature of the electrical energy mix. Moreover, the location of the storage facilities can also be optimized, by minimizing the Joule losses. By using a previously developed model of the French grid, we showed that losses are the lowest when storage devices are installed next to the production facilities, whatever their nature, be it in the present-time mix, or for hypothetical future mixes.

These results yield a generalization of Pellow's EROI formula to the full power grid extent: 
\large
\begin{equation}\label{pelloweroigrid_generalized}
    \text{EROI}_\text{grid} = \frac{(1 -\varphi)\zeta_{\mathrm g\to \mathrm d} + \eta_{\mathrm{st}}  \varphi\zeta_{\mathrm s\to \mathrm d}}{\frac{1}{\text{EROI}_{\text{gen}}} + \frac{\varphi \zeta_{\mathrm g\to \mathrm s}}{\text{ESOI}}+\frac{(1 - \varphi)\zeta_{\mathrm g\to \mathrm d} + \eta_{\mathrm{st}} \varphi\zeta_{\mathrm s\to \mathrm d}}{\text{ETOI}}}
\end{equation}
\normalsize
where $\zeta_{\mathrm g\to \mathrm d}$ refers to the fraction of power remaining after the Joule losses happening between generation and distribution. $\zeta_{\mathrm g\to \mathrm s}$ and $\zeta_{\mathrm s\to \mathrm d}$ are for the Joule losses between generation and storage, and between storage and distribution. In the efficient case where storage is located at the production's maxima, we can approximate $\zeta_{\mathrm s\to \mathrm d}\simeq \zeta_{\mathrm g\to \mathrm d}$ and $\zeta_{\mathrm g\to \mathrm s}\simeq 1$. The new acronym $\text{ETOI}$ stands for Energy Transmitted on energy Invested. It is obtained by the ratio of the final energy transmitted to the consumer through the various layers of power grids by the total amount of energy invested in construction, operation and decommissioning of the network infrastructure. Optimizing storage among the power grid system implies maximizing this $\text{EROI}_\text{grid}$, defined in equation \ref{pelloweroigrid_generalized}. To do so, an optimal sizing strategy aims to find the value of $S_{\text{max}}$ that maximizes the $\text{ESOI}$, an optimal location strategy that maximizes the $\zeta$ factors and an optimal graph topology that maximizes the $\text{ETOI}$. 

This work is subject to several limitations. First, the storage dynamics model assumes a fixed power-to-energy ratio $P_\text{max}/S_\text{max} = 1/\Delta t$, which corresponds to a C-rate of $1\;\text{h}^{-1}$ given the hourly time step. In practice, the embodied energy per unit of stored energy $\zeta_E$ and per unit of peak power $\zeta_P$ do not scale identically when this ratio varies, and the resulting impact on the energy investment $E_I^s$ (Eq.~\ref{storagestock}) remains unexplored. Second, the Haar wavelet filtering restricts the analysis to a single timescale window ($6$--$12\;\text{h}$). Real power systems require storage technologies operating across multiple timescales---from sub-hourly frequency regulation to seasonal balancing---and the interactions between these timescales, in particular how sizing at one timescale constrains or relaxes the requirements at another, are not captured by the present framework. To complete the list of limitations, it is obvious that numerical results and conclusions depend on the precise values of energy intensities considered. Considering that Renewable Energy Technologies benefit from a steady improvement trend of its energy investment figures \cite{steffen_historical_2018}, using updated data would certainly bring finer results.

Even though the methodologies vary, the ESOI-based sizing estimate results obtained in Table \ref{tab:FrenchIndicatorComp} are consistent with the ones reported in Ref.\cite{clerjon_matching_2019}, and compatible with the topological analysis. This methodology using graph centralities and
stored energy fluxes is a new framework that should be applied to different cases of storage deployment, in other countries, group of countries. It would be very profitable to study storage at a local, individual consumer level, while keeping a global accounting eye on the whole power grid. The overall allocation strategy modeling developed in this paper offer new perspective and heuristics on the characterization of future power grids. This model is compatible with the previously published work on spatial topology modeling for power grid \cite{emery_power_2025}. It is highly flexible and can be adapted to the needs of the power grid in question.

\section*{Acknowledgements} We would like to acknowledge interesting discussions with Marc Petit and Jing Dai, as well as feedback from Davide Zanchetta.

\section*{Author Contributions}
HB and SA supervised this work. EE performed the investigation, formal analysis, software development, and initial draft writing. All authors contributed to the study’s conceptualization, validation, final writing, review, and editing.

\section*{Conflicts of interest}
There are no conflicts to declare.

\section*{Data availability}

The data used are open-access \cite{ODReenergymix,ODReinstallations,conso}.

\section*{Code availability}
The code developed for this paper is available at \url{https://doi.org/10.5281/zenodo.18651484}.\\


\balance


\bibliography{biblio} 
\bibliographystyle{rsc} 





\appendix

\section*{Supplementary information}




\begin{figure*}[b]
    \vspace{2.2cm}
    \centering
    \includegraphics[width=\linewidth]{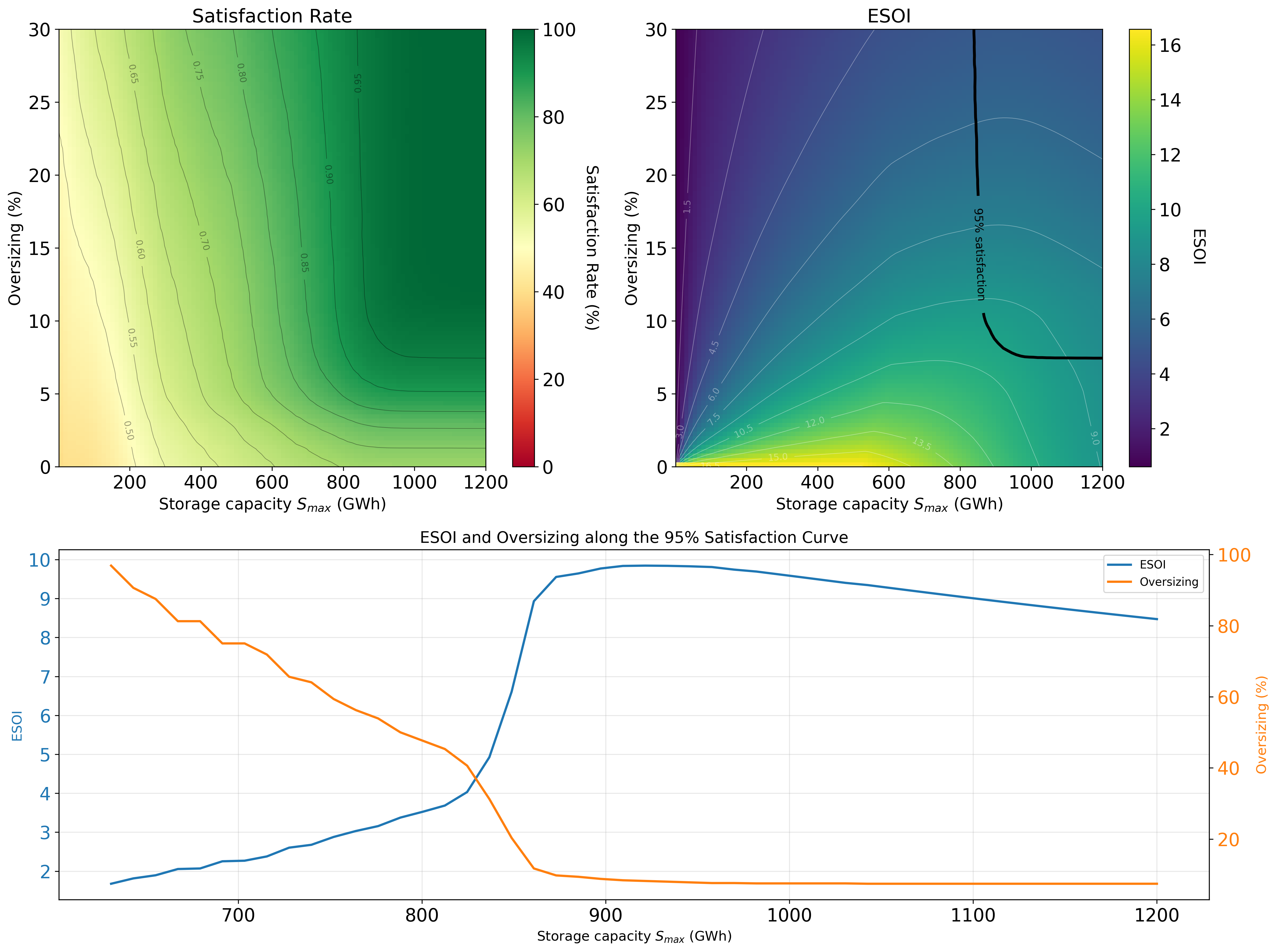}
    \caption{Equivalent of Figure \ref{fig:Smax} but for a 100\% PV energy mix.}
    \label{fig:Smax_pv}
\end{figure*}

\begin{figure*}[h]
    \centering
    \includegraphics[width=\linewidth]{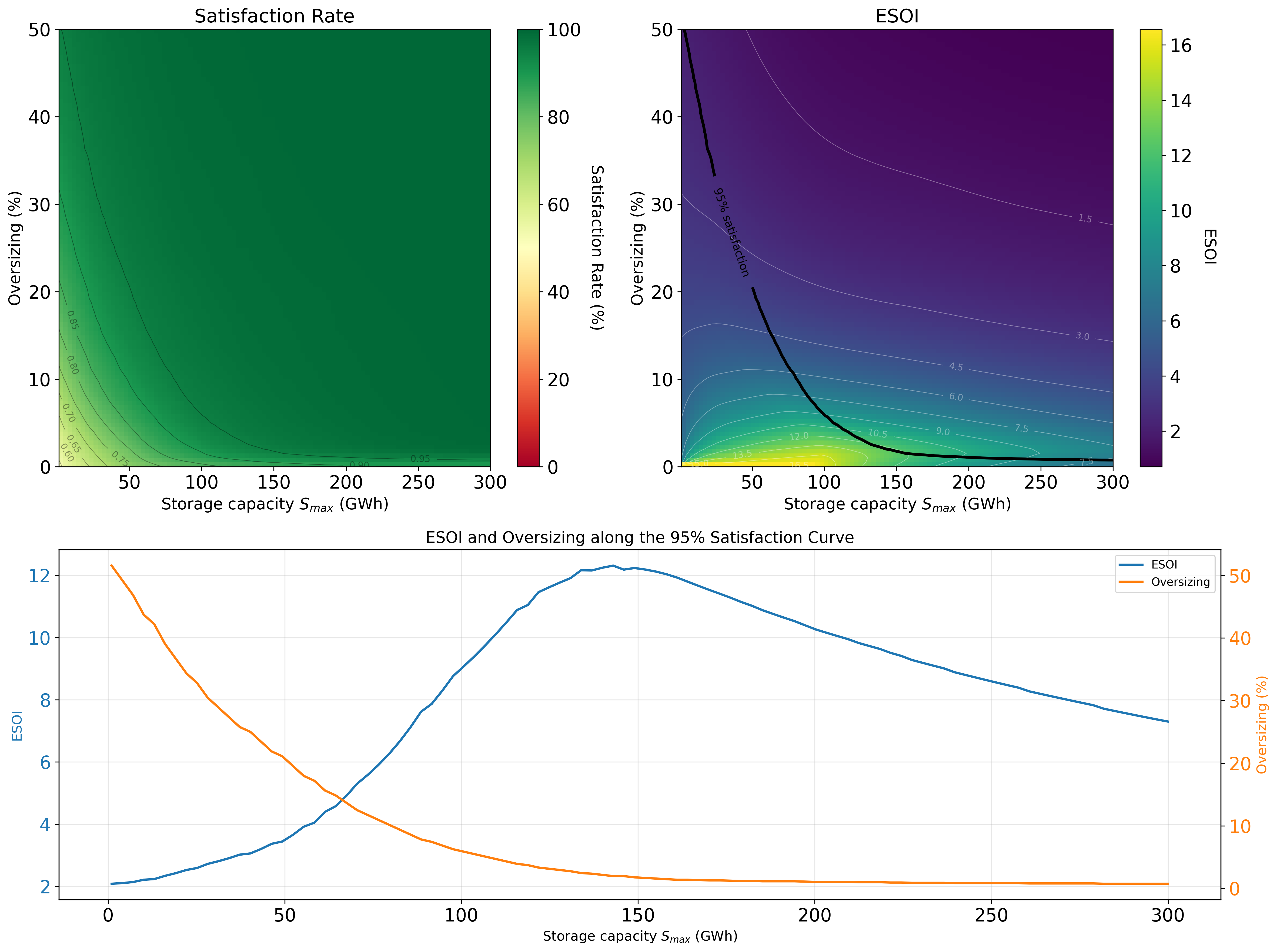}
    \caption{Equivalent of Figure \ref{fig:Smax} but for a 100\% Wind energy mix.}
    \label{fig:Smax_wind}
\end{figure*}

To estimate the optimal storage capacity for other types of energy mix, we applied similar methodologies for a modified input. Here the replacement step from \ref{replacement} is more radical. All production signals are fixed at zero except the signal of interest which rescaled proportionally to the removed production. We use the 100\% PV signal ${P}(t)=\alpha_\text{S}P_{\mathrm{S}}(t)$ in Figure \ref{fig:Smax_pv} and the 100\% Wind signal ${P}(t)=\alpha_\text{W}P_{\mathrm{W}}(t)$ in Figure \ref{fig:Smax_wind}. The strong non-linear variation ocuring around  $S_\text{max}^* \sim 850$ GWh of capacity in the figure \ref{fig:Smax_pv} could be explained as follows. Before $S_\text{max}^*$ the variation are such that satisfaction increase through direct additional storage capacity: the production peak must be stored. After $S_\text{max}^*$, the production peaks are stored but the over-consumption periods that cannot be fed by pure storage remains: the solution to this is not storage but oversizing.

\end{document}